\documentclass[twocolumn,caps,showpacs]{revtex4}

\usepackage{graphicx,color}
\usepackage{latexsym}
\usepackage{amsmath,amssymb}
\usepackage[draft=false]{hyperref}
\usepackage{mathrsfs}
\usepackage{comment}
\usepackage{mathrsfs}
\usepackage{subfigure}
\DeclareSymbolFontAlphabet{\mathrsfs}{rsfs}
\DeclareMathAlphabet{\mathcal}{OMS}{cmsy}{m}{n}

\begin{document}

\title{Estimation of Reynolds number for flows around cylinders with lattice Boltzmann methods and artificial neural networks}

\author{Mauricio Carrillo}
\affiliation{Instituto de F\'{\i}sica y Matem\'{a}ticas, Universidad
	Michoacana de San Nicol\'as de Hidalgo. Edificio C-3, Cd.
	Universitaria, 58040 Morelia, Michoac\'{a}n,
	M\'{e}xico.}

\author{Ulices Que}
\affiliation{Instituto de F\'{\i}sica y Matem\'{a}ticas, Universidad
	Michoacana de San Nicol\'as de Hidalgo. Edificio C-3, Cd.
	Universitaria, 58040 Morelia, Michoac\'{a}n,
	M\'{e}xico.}
	
\author{Jos\'e A. Gonz\'alez}
\affiliation{Instituto de F\'{\i}sica y Matem\'{a}ticas, Universidad
	Michoacana de San Nicol\'as de Hidalgo. Edificio C-3, Cd.
	Universitaria, 58040 Morelia, Michoac\'{a}n,
	M\'{e}xico.}

\date{\today}

\begin{abstract}
The present work investigates the application of Artificial Neural Networks (ANNs) to estimate the Reynolds ($Re$) number for flows around a cylinder. The data required to train the ANN was generated with our own implementation of a Lattice Boltzmann Method (LBM) code performing simulations of a 2-dimensional flow around a cylinder. As results of the simulations, we obtain the velocity field ($\vec{v}$) and the vorticity ($\vec{\nabla}\times\vec{v}$) of the fluid for 120 different values of $Re$ measured at different distances from the obstacle and use them to teach the ANN to predict the $Re$. The results predicted by the networks show good accuracy with errors of less than $4\%$ in all the studied cases. One of the possible applications of this method is the development of an efficient tool to characterize a blocked flowing pipe.
\end{abstract}


\maketitle

\section{\label{sec:Int} Introduction}
The analysis of  incompressible fluid flows around obstacles has applications to many relevant physical problems including the study of the mechanical properties of foams \cite{Dollet}, aerodynamics of bridges \cite{Larsen}, wave patterns generated in the atmospheric flow \cite{Lacaze} among others. These flows have a transcendental importance because they are present in a great variety of fields ranging from chemical and industrial engineering to environmental, physical and biological processes. The study of those scenarios could be difficult due to the boundary layer effects and recirculation zones present in such flows, for which viscous forces are either dominant or comparable with inertial forces. Predictions of the fundamental properties of fluids are a difficult task, especially in the case of industrial applications which require high precision for complex systems.

Given its singular complexity, simulation of flows around an obstacle are one of the most common problems of study in computational fluid dynamics (CFD). In the last decades grid-based numerical methods such as finite element method \cite{Tuann}, and finite volume method \cite{Rahman} were commonly used to simulate these flows. Other methods such as smoothed particles hydrodynamics \cite{Monaghan} and Lattice Boltzmann Method (LBM)  \cite{Mohamad} have also become popular due to the good results they have shown performing these simulations, together with the considerable simplicity of their implementation.

We are interested in a specific application of the phenomenon of flows around obstacles, which is of great importance for hydraulic engineering, for instance in the prediction of the shape and location of objects blocking the flow in a pipe \cite{Lee, Wang}. Therefore, as an initial step we have developed a LBM numerical code introducing the physical properties of the problem such diameter and location of the obstacle, density, viscosity and initial velocity of the fluid to simulate the flow around a cylinder, obtaining physical information such velocity, vorticity or density of the fluid along the domain. After the simulations are made, we extracted the $x$-component of the velocity field ($v_x$) and the $z$-component of the vorticity for fluids with Reynolds number ($Re$) between 1 and 120. In this work, we implemented a machine learning (ML) method with a supervised training algorithm, which uses the velocity field or vorticity as input information, and as target a physical property characterizing the flow, in this case the corresponding $Re$ for the input data used at each simulation. After training, the only information needed are the velocity or vorticity profiles to predict the $Re$. This method, is proposed as the first step towards a more general algorithm capable to characterize physical properties of objects blocking flow pipes. The ML method employed uses Artificial Neural Networks (ANNs), which are considered because they offer a direct method to solve problems given their adaptation to unknown, incomplete or noisy information besides a fast computation after training \cite{Rabunal}. In addition, they have been successfully applied in other complex systems  like image and voice processing, pattern recognition, signals filtering and also have been implemented in computational fluid modeling for turbulent flows and dispersion \cite{Lauret2015,Lauret2014,Lauret2013} which is pretty similar to the vector field patterns analyzed here.

The paper is organized as follows: section \ref{sec:FSLBM} is a detailed description of the problem of a flow around a cylinder, the Lattice Boltzmann method and the tools employed to perform the fluid's simulations, section \ref{sec:ANNs} describes ANNs concepts, in section \ref{sec:Data} we present the process applied to the data obtained from the simulations in order to be introduced into the network, section \ref{sec:Results} describes the results and finally in section \ref{sec:Conclusions} we present the conclusion.

\section{\label{sec:FSLBM} Simulating a 2-Dimensional flow with LBM}

The numerical simulation of the flow around  a cylinder has been widely investigated in numerous papers such as \cite{Catalano, Grucelski, Karabelas}. The most common scenario  is to consider the obstacle (the cylinder) immersed in an infinite medium (free flow). In the case of an incompressible viscous fluid, the pattern of this flow varies depending on the Reynolds number, and it is common to express its physical significance as the ratio between the inertial and viscous forces.

\begin{figure}
\includegraphics[width=9cm]{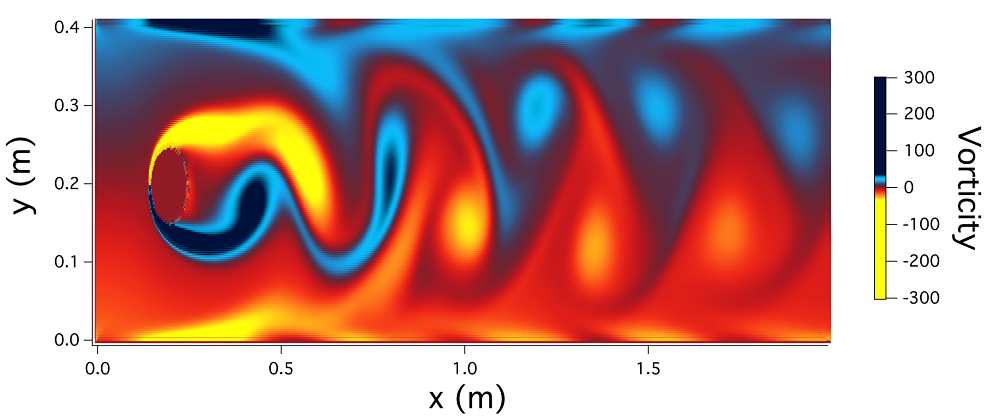}
\caption{A Karman vortex street is a sequence of alternated rotating vortices caused by the unsteady separation of the boundary layer of a flow passing over submerged bodies. This plot was obtained using our own LBM code for a $Re = 100$.}\label{fig:Karman}
\end{figure}

For small Reynolds numbers ($Re<10$) the fluid pressure rises from the free stream value to the stagnation point. As the $Re$ increases, the high pressure constrains the fluid to move along the cylinder surface, generating boundary layers that separate the flow on both sides. At the same time shear layers are generated next to the cylinder, at the top and bottom of the flow. These layers are rolled on themselves generating vortices rotating in opposite directions relative to each other. As $Re$ exceeds the value of 40, the wake after the cylinder becomes unstable and the known Karman vortex street is generated as the result of an alternated projection of vortices recurrently(see Figure \ref{fig:Karman}).

To obtain a numerical prediction, we construct a numerical code based on LBM for the 2 dimensional case of the flow around a cylinder. LBM is a numerical procedure to study CFD with an increasingly popularity because of its simplicity of implementation and high capacity to simulate a wide variety of phenomena \cite{Aidun, Sheikholeslami}.

Let us present a small description of the method, focusing on the 2-dimensional flow around cylinder (for more details see \cite{Mohamad, Guo, Chen}). The fundamental idea is that fluids can be visualized as a large number of small elements moving with random motions, through a lattice of cells in nine possible directions. One of these directions represents no motion of the fluid at all and the other eight represent the motion in the plane with angles of $45^{\circ}$ between each other. In each of these cells the physical quantities that represent the fluid, like the density and the pressure for example,  are calculated. The elements representing the fluid are considered like ensembles of particles described by a distribution functions $f$. The exchange of momentum and energy in the fluid happens through particle streaming and particle collision and a simple discretization of the Boltzmann equation given by
\begin{equation}
\label{ecLB1}
f_i(x+c_i\delta t, t + \delta t) = f_i(x,t) + \frac{1}{\tau}\left(f_i^{eq}(x,t) - f_i(x,t)\right),
\end{equation}
\noindent where the index $i$ represents the $i-$th discrete lattice direction, $c_i$ is the streaming velocity, $f_i^{eq}$ the equilibrium distribution function, and $\tau$ the relaxation time for $f_i$ needed to reach $f_i^{eq}$. He and Luo in \cite{He} have shown that this expression can be derived from the continuous Boltzmann equation. The last term in Eq. (\ref{ecLB1}) performs the collisions describing the variation of the number of particles moving in each direction on the lattice due to microscopic inter-particle collisions.

The equilibrium distribution function $f_i^{eq} (x, t)$ (whose derivation can be seen in \cite{Mohamad} and \cite{He}) is used to determine the local velocity of fluid elements due to collisions, and is computed through the macroscopic variables so that mass, momentum and energy are preserved for each cell and it is expressed as
\begin{equation}
f_i^{eq}(\rho, \vec{ v}) = \rho w_i \left(1+\frac{3}{c^2}\vec{ c}_i \cdot \vec{ v} + \frac{9}{c^4}(\vec{ c}_i \cdot \vec{ v})^2 -\frac{3}{2c^2}\vec{ v}^2\right),
\end{equation}
\noindent with $\rho$ the density, $\vec{ v}$ the velocity, $c$ the ratio between the lattice grid spacing and the time step, and $w_i$ a weighting factor.

Macroscopic variables such as $\rho$ and $\vec{ v}$ are computed through the distribution functions $f_i$ in the particle velocity space as
\begin{equation}
\rho(x,t)= \sum_i f_i^{eq}, \;\;\;\;\;\;  \vec{ v}(x,t)= \frac{1}{\rho}\sum_i f_i^{eq}\vec{ c}_i,
\end{equation}
\noindent while pressure $p$ is computed from an equation of state.

For a computational implementation, the external boundary conditions are imposed at a finite distance but far enough such that the characteristic flow parameters are not affected by the internal calculations. To implement the boundary representing the income fluid (at the left of the numerical domain), we used the model developed by Zou and He in \cite{Zou}, applying a velocity boundary using a given velocity profile as input for the numerical modeling. In turn, for the boundary where the fluid exits (at the right of the domain), we use the method proposed by Yu et al. \cite{Yu}.

We considered a flow moving in a direction oriented in the positive $x$ axis with a cylindrical obstacle of a diameter $l_c=0.1$m fixed near the left boundary of the domain. The domain of the simulation has a total length of $L_y = 0.41$m in the vertical direction and $L_x=2$m along the horizontal. Density and kinematic viscosity are chosen as $\rho=10^{3}\rm{kg}/ \rm{m}^3$  and $\nu=10^{-3}\rm{m}^2/\rm{s}$ respectively. Accordingly, given a certain characteristic velocity $v_c$, we can calculate the $Re$
\begin{equation}
\label{ecRe}
Re = \frac{v_c l_c}{\nu},
\end{equation}
\noindent where $l_c$ is the characteristic length of the system represented in this case by the diameter of the cylinder. The initial velocity profile at the inlet of the domain of the system is used as input in the simulation, we use a Poiseuille flow determined by
\begin{equation}
\label{PoiExactVmax}
v_x(y) = 6v_{c}\frac{y(l_c-y)}{l_c^{2}},
\end{equation}
\noindent where $v_{c}$ is the characteristic velocity of the flow which is equal to 2/3 of the maximun velocity of a stationary solution for the velocity field.

\begin{figure}
\includegraphics[width=9.0cm]{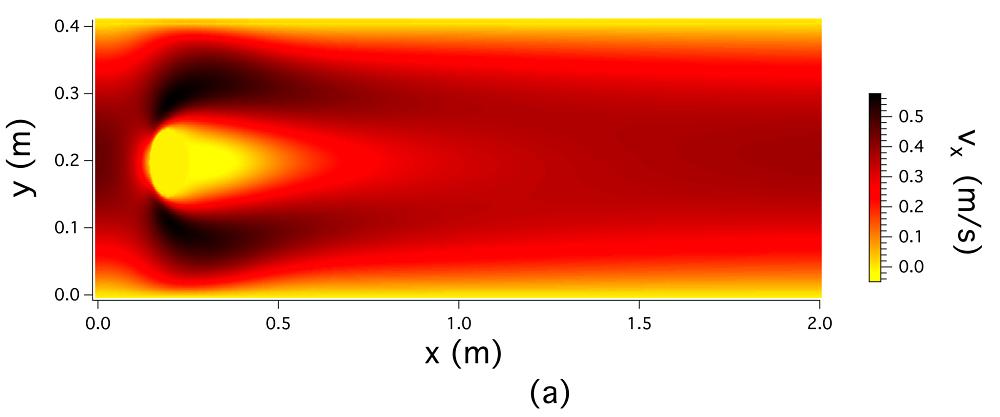}
\includegraphics[width=9.0cm]{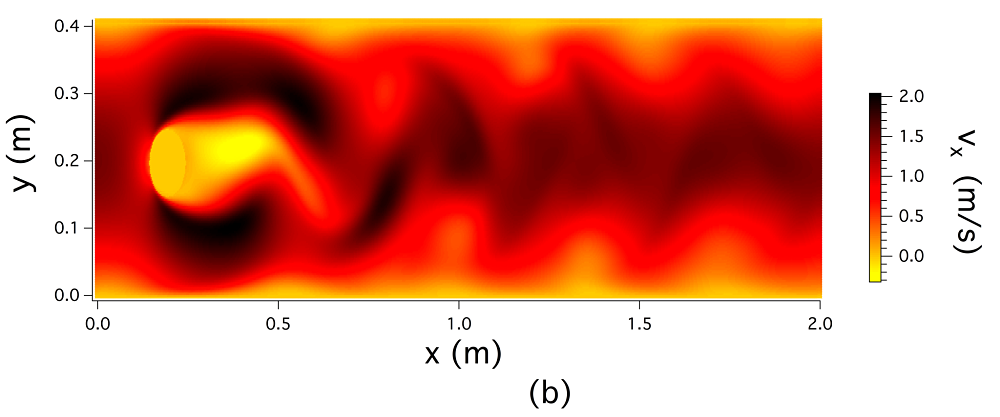}
\caption{Magnitude of $v_x$ for the flow around a cylinder with $Re=30$ (a) and $Re=100$ (b). It is noted that the vortices formed at the cylinder at low Reynolds disappear while for high Reynolds generate the characteristic Karman vortex street.}
\label{fig:FACVx}
\end{figure}

To build the database necessary to train the neural network, several numerical simulations of the flow around a cylinder were performed using the LBM described above. Simulations for 120 different values of $Re$ were generated starting from $ Re=1$ to $Re=120$ in steps of $\Delta Re = 1$.  A mesh of $164\times820$ cells was used and the only free input parameter for each simulation was the inlet velocity profile parametrized by $v_c$. We increase the resolution of the domain of the simulation to prove the convergence of the results for the numerical simulation. We found that the physical values obtained for the increased resolution (duplicated to 328 cells in the y axis) are very close to those obtained originally. The functions $v_x$ and the $z$-component of the vorticity were measured when the neutral stability was reached as shown in Figure \ref{fig:FACVx}. It is noteworthy that the phenomenon of the von Karman vortex street is clearly seen in the results. The way this information is entered into the ANNs is explained in section \ref{sec:Data}, but before that a brief introduction on ANNs is in order.

\section{\label{sec:ANNs}Artificial Neural Networks concepts}
ANNs are a ML paradigm based on biological systems used for classification, optimization and regression problems, using a massive number of interconnections of nodes or neurons, in reference to biological neural networks \cite{Russell}. ANNs try to generalize the most relevant information on noisy or incomplete data, extracting patterns through the composition of nonlinear functions (usually sigmoids like the logistic or hyperbolic tangent), working as universal function approximators, like the theorem of Cybenko states \cite{Cybenko}.

The most usual ANN architecture is the so called multilayer perceptron (MLP), which is an extension of the simple Perceptron developed by Rosenblatt \cite{Rosenblatt}. The MLP consists of an arrangement of layers of processing neurons interconnected between them: input layer , one or several hidden layers and an output layer. In a MLP structure information goes from the input to the hidden layer and later from the hidden to the output one. For neurons in hidden and output layers, the connections are regulated by weight coefficients, which are adjusted in the phase called training.

To illustrate an ANN operation, suppose a three layer MLP with $n$ input neurons, $m$ elements in a single hidden layer and $l$ output neurons, where the relation between an input vector $\vec{x} = \{x_1,x_2,\dots,x_n\}$ and the $k$-th output neuron is determined by the expression
\begin{equation}
	y_k = G\left(\tilde{w}_{0k} + \sum_{j=1}^{m} \tilde{w}_{jk} * F\left(\sum_{i=1}^{n} w_{ij} * x_i + w_{0j} \right) \right),
	\label{eq:ANN_output}
\end{equation}
\noindent where $w_{ij}$ and $\tilde{w}_{jk}$ are the connection weights between the neurons in input-hidden and hidden-output layers respectively; $w_{0j}$ and $\tilde{w}_{0k}$ are some extra weights called bias operating as thresholds; $F$ and $G$ are the neurons' activation functions on each layer, in this case we set $F$ as the hyperbolic tangent and $G$ a linear function.

In order to ensure that an ANN gives proper results, its weights must be adjusted. One way to do this is implementing a supervised training algorithm, which consists in introducing $N$ examples into the ANN. Each example consists in $i$ inputs denoted by $\tilde{x}_{i}^p$ and $k$ outputs denoted by $\tilde{y}_{k}^p$, where the index $p=1,\dots,N$. A minimization of the weight dependent error function $E(\vec{w})$ defined by
\begin{equation}
E(\vec{w})=\frac{1}{N} \sum_{p=1}^{N} \sum_{k=1}^{l} \frac{1}{2}(\tilde{y}_k^p - y_k^p)^2,
\end{equation}
is performed. In this paper $\tilde{x}_i^p$ corresponds to the $i$ values measured for $v_x$ or the $z$-component of the vorticity at the detectors for the $p-$th simulation, while $\tilde{y}_k^p$ represents the corresponding $Re$ value for the same simulation (having only one output in the network implies $k=1$). The minimization of the cost function is made by the iteration of a gradient descent algorithm called \textit{backpropagation} \cite{Rumelhart}, which searches the direction in which the error decreases the most on each step, and then changing the value for each $w_{ij}$ in the next time step $t+1$, propagating the error at time $t$ from output to input neurons by the rule
\begin{equation}
w_{ij}(t+1)= w_{ij}(t) - \gamma \frac{\partial{E(t,\vec{w})}}{\partial{w_{ij}(t)}} + \alpha \triangle w_{ij}(t),
\end{equation}
\noindent with $0<\gamma<1$ called the learning rate, determined by the user, $\alpha \triangle w_{ij}(t)$ is called the momentum term, added for preventing getting trap in a local minimum,  and $0<\alpha<1$ another parameter to be adjusted.

In this work the MLP neural network has been chosen for its easy implementation, however there are more sophisticated ANNs architectures which could improve the results or suitable in more complex scenarios, for example, considering the time evolution of the flow patterns. And as a first step towards this problem, we also inspect the ANN predictions for flow patterns at different times, as will be described later. In addition, given that the results for a single ANN might be different from another ANN, either because the weights are randomly initialized, or the selected learning rate and/or momentum values, could improve or decrease their prediction accuracy, even with the same number of iterations or the selection of training, validation and test sets. The predictions presented in the following sections are the average of ten different ANNs outcomes, and representative for using ANNs with particular parameter values defined in section \ref{sec:Results}.

\section{\label{sec:Data} Data processing}
\begin{figure}
\includegraphics[width=0.45\textwidth]{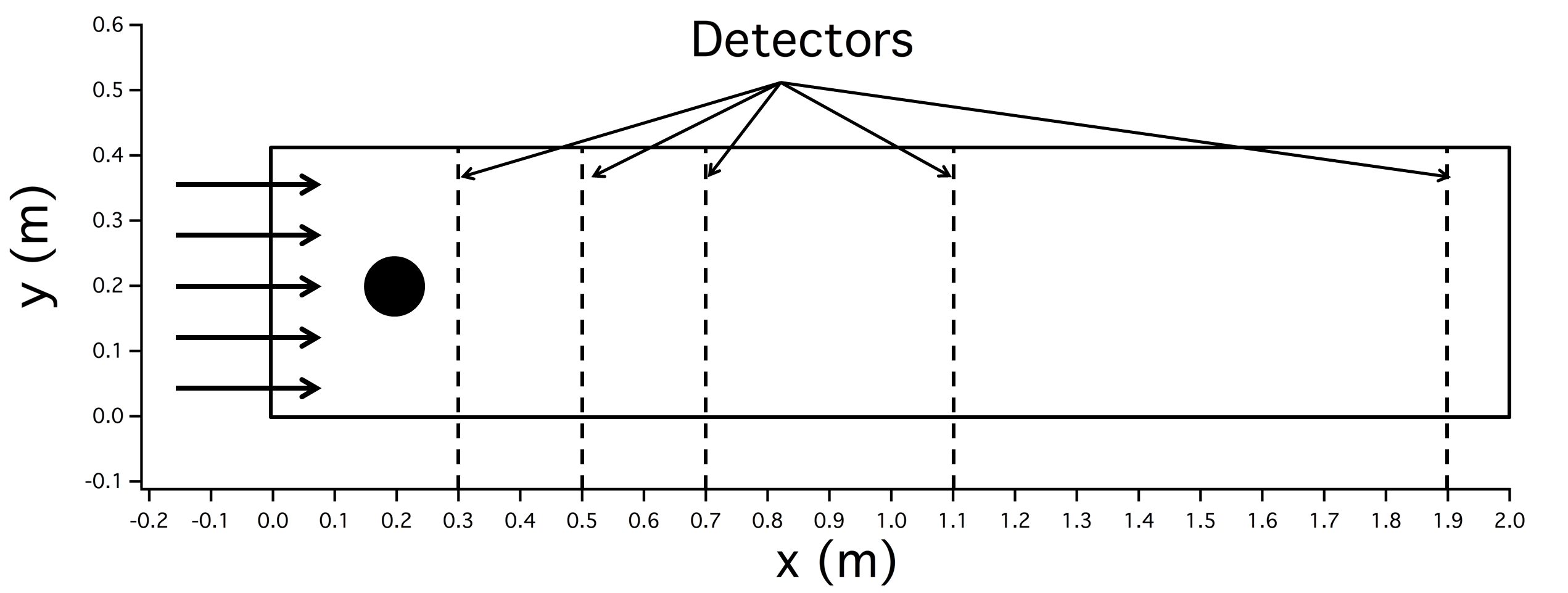}
%
\caption{A schematic representation of the flow around the cylinder and locations of the detectors where 1,4,6,11,21,41 and 82 values of $v_x$ and the $z$-component of the vorticity were extracted. The fluid moves from left to right and the measurement is performed when the neutral stability is reached. The times vary from $t=20$s to $t=50$s depending on the value of $Re$.}
\label{fig:Schematics}
\end{figure}

\begin{figure}
\includegraphics[width=9.0cm]{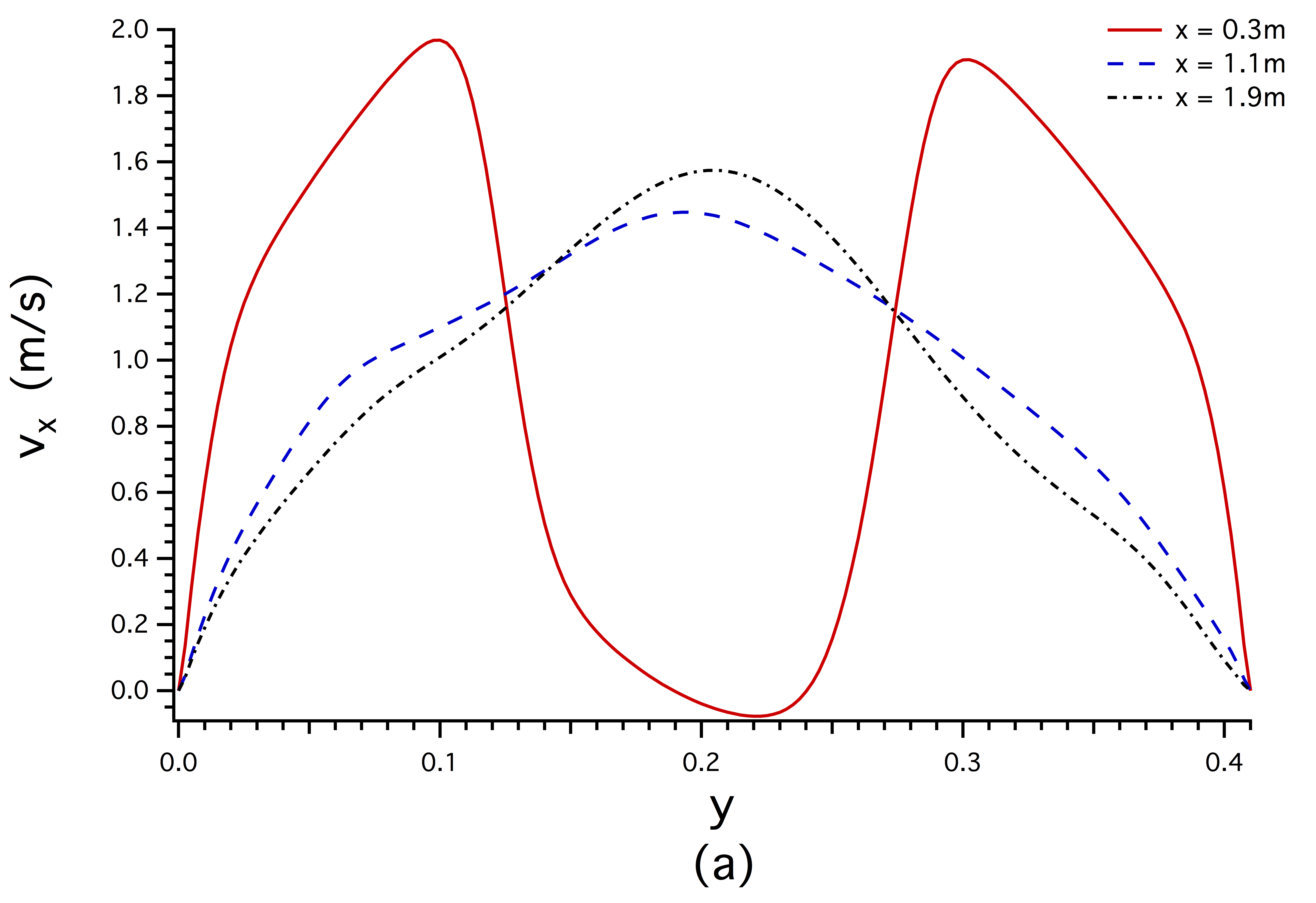}
\includegraphics[width=9.0cm]{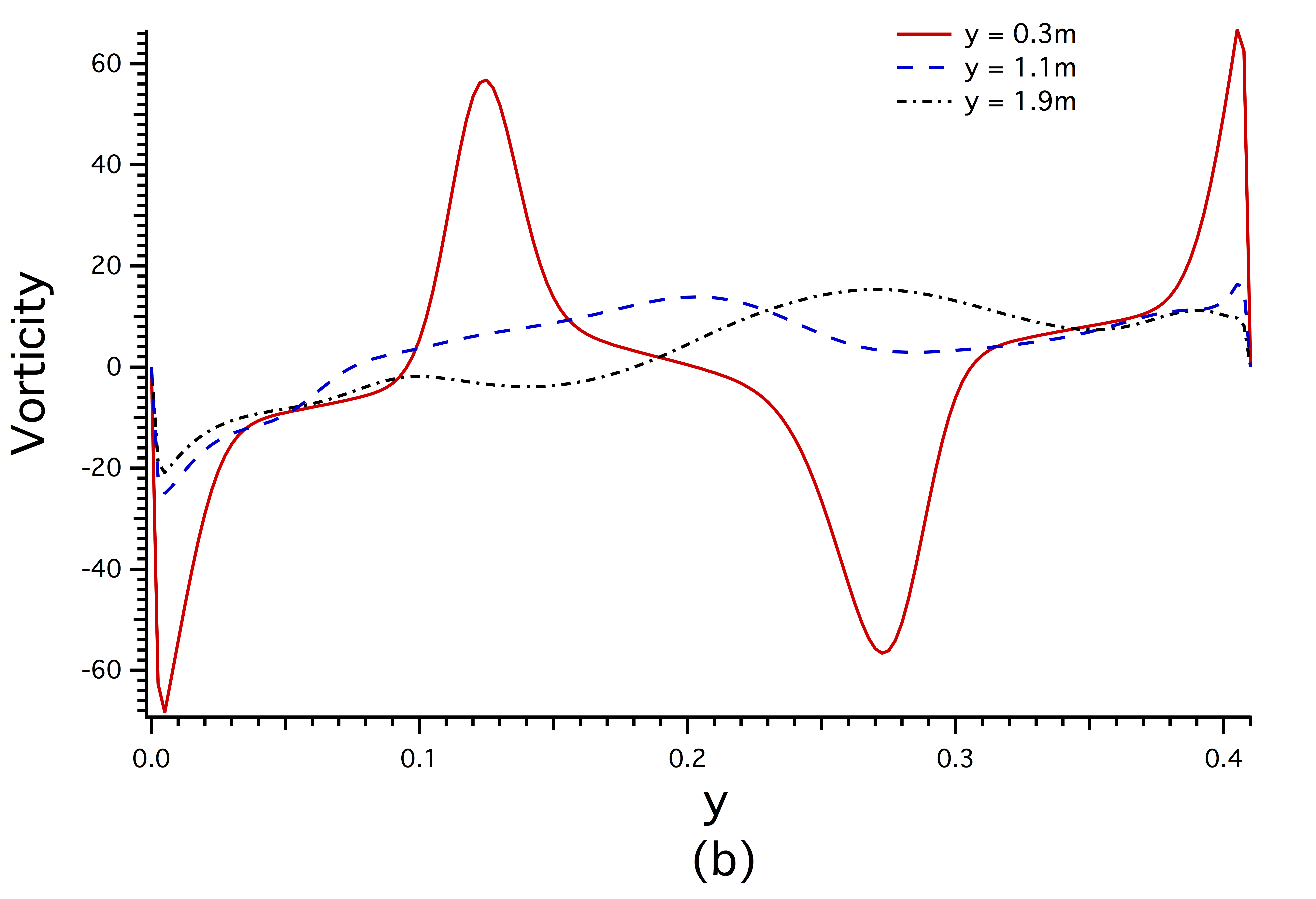}
\caption{Example of the measurements obtained for different locations of the detectors for $Re=100$. We plot the $x$-component of the velocity field (a) and the $z$-component of the vorticity (b) versus the position in $y$, for detectors at $0.3$m, $1.1$m and $1.9$m over the $x$ axis.}
\label{fig:Inputs_x}
\end{figure}

\begin{figure}
\includegraphics[width=9.0cm]{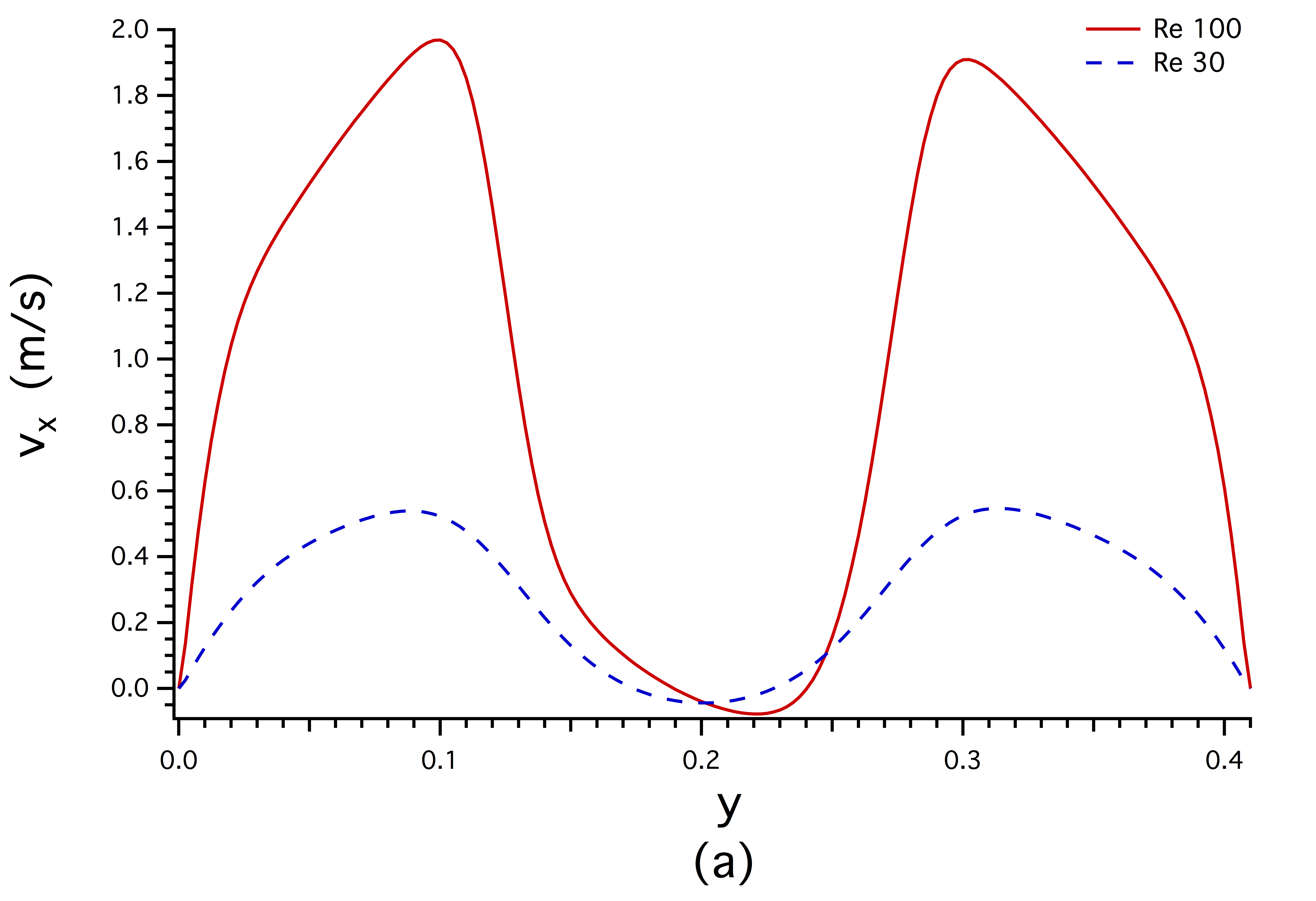}
\includegraphics[width=9.0cm]{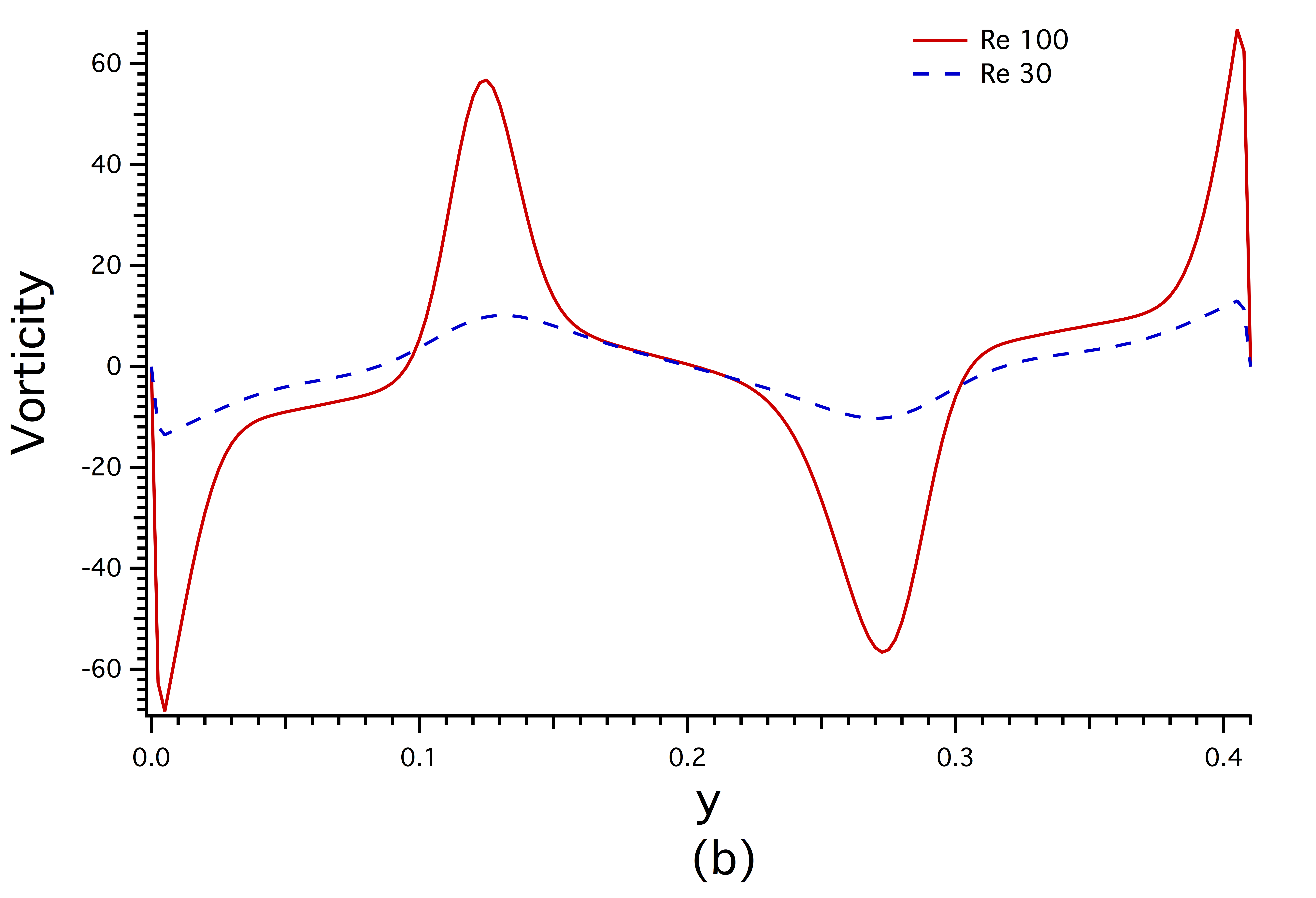}
\caption{Same as Figure \ref{fig:Inputs_x} but in this case we fix the detector at $x=0.3$m and change the value of the Reynolds number. In (a) is presented the $x$-component of the velocity field and at (b) the $z$-component of the vorticity.}
\label{fig:Inputs_Re}
\end{figure}

In order to predict the Reynolds number, we extract physical information from the fluids' velocity component along the $x$-axis and the $z$-component of the vorticity at five locations in the posterior region of the cylinder at $0.3$m, $0.5$m, $0.7$m, $1.1$m and $1.9$m as shown in Figure \ref{fig:Schematics}. The first detector is located immediately behind the cylinder in order to evaluate the predictive capability of the ANN in the area where the vortices are generated. The three intermediate locations represent measurements in regions where one is observing processes from the generation of vortices in the Karman vortex street. Finally, the last detector measures the performance of the ANN when one moves considerably away from the obstacle. Examples of the profile velocity $v_x$ and the vorticity $(\vec{\nabla}\times\vec{v})_z$ for different positions of the detectors and a fixed $Re=100$ are shown in Figure \ref{fig:Inputs_x}, while in Figure \ref{fig:Inputs_Re} the profiles are shown for a fixed detector at $x=0.3$m and different values of $Re$. 

The flow patterns for high $Re$ generated in the simulations are changing recurrently on time as explained in section \ref{sec:FSLBM}. For simplicity, the data is extracted at fixed time after the flow has reached a neutral stability. This fixed time is set for the the highest Reynolds number ($Re = 120$) so that we can extract the data always at the same physical time for all the $Re$ studied, and is approximately equal to $36.5$s or $70,000$ computer iterations. In order to study the dependence of the results on the extraction time, on the next section we study the ANNs predictions at twenty different times (separated by intervals of $2.6$s) for low ($Re=30$) and high ($Re=99$) Reynolds numbers.

On each one of this detectors, a reduction of the number of spatial nodes has been done selecting only half of them in the $y$ direction of the lattice, reducing the number of points where the vector fields are measured from 164 to 82. In addition, we also studied how the ANNs performance is affected if this number is decreased, reducing the number of points to a total of 1,4,6,11,21 and 41. On each case, the measuring points are equidistant starting from the edges, except for the case of a single point located at the center of the detector.

The ANNs constructed for each case have the same number of inputs as the values of the vector field extracted in each detector, i.e., from 1 to 82 inputs. 10 hidden neurons with a hyperbolic tangent as activation function and a single output with a linear function restricted to positive values only since we might not know an upper limit for $Re$. For example, an ANNs' $Re$ prediction following equation (\ref{eq:ANN_output}) using 82 $v_x$ inputs will be defined as
\begin{equation}
Re = \tilde{w}_{0k} + \sum_{j=1}^{10} \tilde{w}_{jk} * \tanh \left(\sum_{i=1}^{82} w_{ij} * v_{x_i} + w_{0j} \right).
\end{equation}

Besides, as is usual on machine learning methods data is split in three different sets, namely, training, validation or test data and prediction set, the latter ones are unknown for the ANN in the weight adjustment process and the backpropagation algorithm is iterated until the error in the test set begins to increase due to overfitting. In this work, simulations were split for training, test and prediction as:
\begin{enumerate}
\item Training: 80 simulations  with $Re$ scattered in a range from 1 to 116.
\item Validation: 20 simulations, different from training set, starting from $Re$ 6 to 117.
\item Prediction: 20 simulations, different from both training and validation sets, with $Re =$ 12, 17, 22, 27, 32, 37, 52, 64, 70, 76, 92, 99, 102, 107, 112, 116,117,118, 119 and 120. The first 15 are in the interpolation zone while the last 5 are in the outer range of training and validation, to test its extrapolation performance, it is expected that the prediction performance decreases far from the interpolation zone.
\end{enumerate}


\section{\label{sec:Results} Results}

A learning and momentum values of $\gamma=0.05$ and $\alpha=0.5$ were used respectively on each ANN, training them until the validation error starts increasing and keeps this behavior for more than 500 iterations, with no particular number of iterations by default. With these parameters the training phase took less than three minutes and from 1000-2000 iterations in general while the time needed to generate the prediction is negligible.

\begin{figure}
\includegraphics[width=9.0cm]{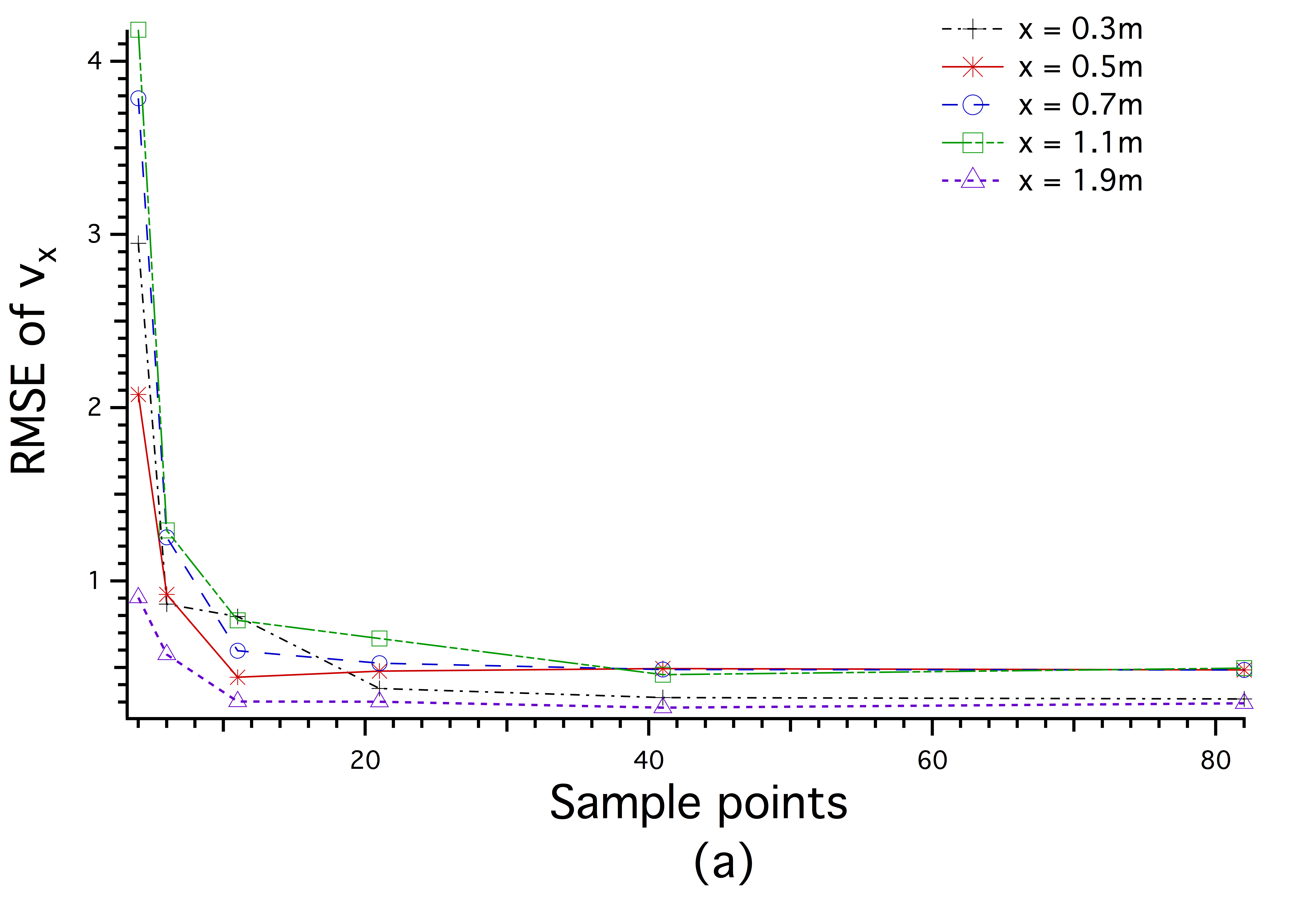}
\includegraphics[width=9.0cm]{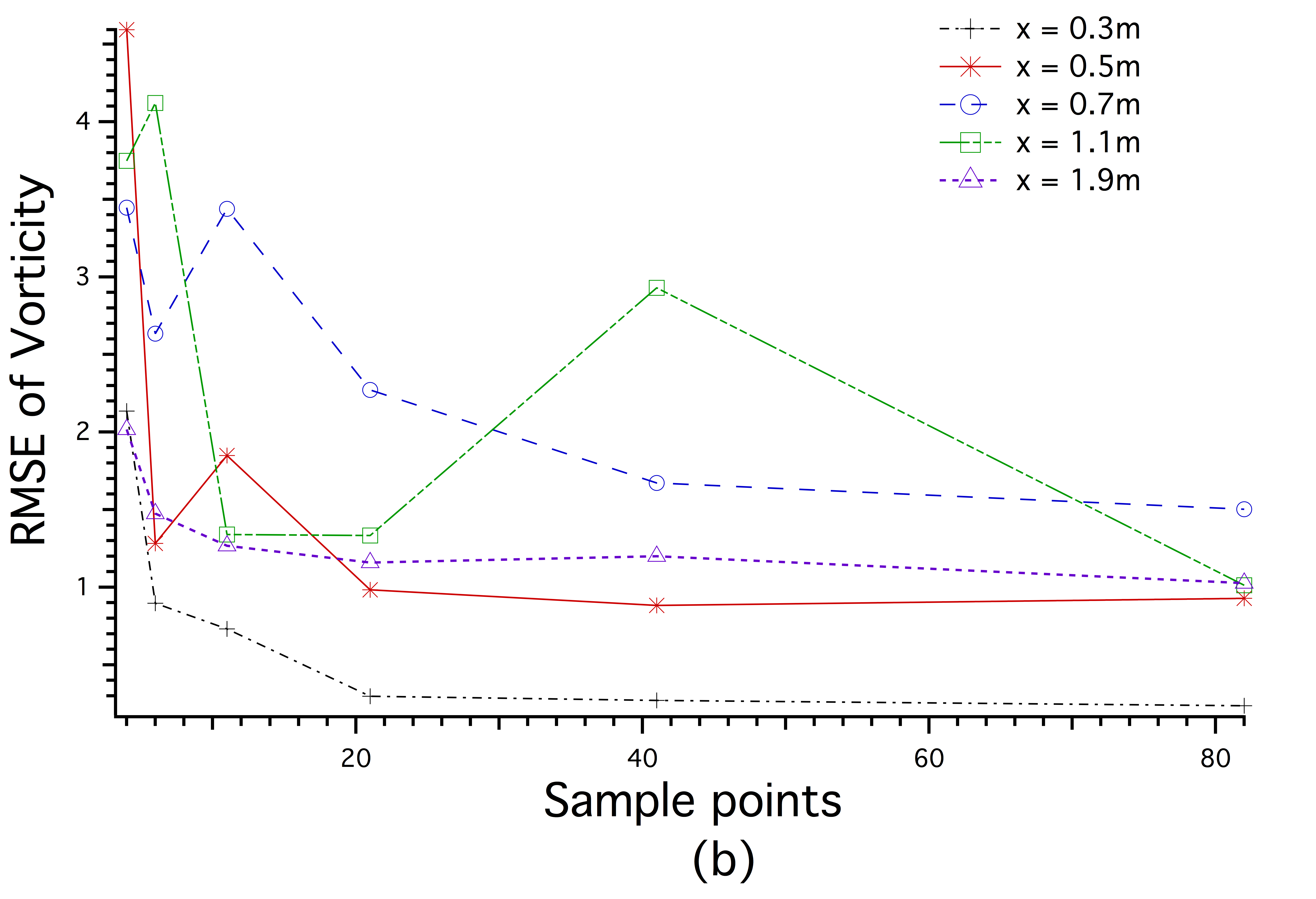}	
\caption{RMSE for the average of the prediction set, measuring at $0.3$m, $0.5$m, $0.7$m, $1.1$m and $1.9$m using $4, 6, 11, 21, 41$ and 82 sampling points using $v_x$ in (a) and the $z$-component of the vorticity in (b). In the plots the RMSE using a single point as input in the ANNs is not considered since that case is out of scale, see Tables  \ref{tab:RMSE_Velx} and \ref{tab:RMSE_Vort}.}
\label{fig:RMSE_velx_vort}
\end{figure}

In Tables \ref{tab:RMSE_Velx} and \ref{tab:RMSE_Vort}, are shown the root-mean-square errors (RMSE) of the average of the predictions for ten trained ANNs on each considered location and for all the cases in the prediction set. On one hand, when considering the values of $v_x(y)$ as the inputs, we observe how the error gets bigger as the number of sampling points is lowered below 11 sampling points, also shown in the top plot of Figure \ref{fig:RMSE_velx_vort}. On the other hand, considering the $z$-component of the vorticity values as inputs, the errors are bigger than using $v_x$. With more than 41 sampling points the RMSE are more alike as shown in the bottom plot of Figure \ref{fig:RMSE_velx_vort}. In fact, we conducted studies using a larger number of adjacent nodes over the $x$ axis without getting a significant increase in accuracy.

\begin{table}
	\centering
	\begin{tabular}{|c|c|c|c|c|c|}\hline
		\multicolumn{6}{|c|}{RMSE using $v_x$} \\ \hline \hline
			{} & 0.3m & 0.5m & 0.7m & 1.1m & 1.9m \\ \hline
		
		1 &	36.652 & 43.596 & 9.5976 & 3.931 &	3.446 \\ \hline
		4 &	1.950 & 2.077 & 3.786 &	1.880	& 0.904 \\ \hline
		6 &	0.421 & 0.921 & 1.252	& 0.528 & 0.575 \\ \hline
		11 & 0.366 & 0.443 & 0.597 & 0.370 & 0.304 \\ \hline
		21 & 0.141 & 0.479 & 0.525 & 0.228	& 0.302 \\ \hline
		41 & 0.133 & 0.494 & 0.488 & 0.227 & 0.268 \\ \hline
		82 & 0.132 & 0.486 & 0.486 & 0.220 & 0.294 \\ \hline
		
	\end{tabular}
	\caption{RMSE of the average prediction at measurement points along the $x$-axis at $0.3$m, $0.5$m ,$0.7$m, $1.1$m and $1.9$m; using $v_x$, obtained with $1, 4, 6, 11, 21, 41$ and 82 sampling points. Observe how the RMSE increases dramatically taking less than 6 sample points.}
	\label{tab:RMSE_Velx}
\end{table}

\begin{table}
	\centering
	\begin{tabular}{|c|c|c|c|c|c|}\hline
		\multicolumn{6}{|c|}{RMSE using vorticity} \\ \hline \hline
		{} & 0.3m & 0.5 & 0.7m & 1.1m & 1.9m \\ \hline
		\text{1} & 42.625 & 26.614 & 30.177 & 21.029 &	22.663 \\ \hline
		4 & 2.135 & 4.593 & 3.447 & 3.748 & 2.016 \\ \hline
		6 & 0.897 & 1.282 & 2.634 &	4.120 &	1.474 \\ \hline
		11 & 0.731 & 1.850 & 3.439 & 1.340 & 1.267 \\ \hline
		21 & 0.298 & 0.983 & 2.271 & 1.333 & 1.159 \\ \hline
		41 & 0.271 & 0.883  & 1.671 & 2.930 & 1.200 \\ \hline
		82 & 0.236 &  0.928 & 1.503 & 1.012 & 1.026 \\ \hline
	\end{tabular}
	\caption{Same as in Table \ref{tab:RMSE_Velx} but using the $z$-component of the vorticity instead of $v_x$.}
	\label{tab:RMSE_Vort}
\end{table}

\begin{figure}[p]
\includegraphics[width=9.0cm]{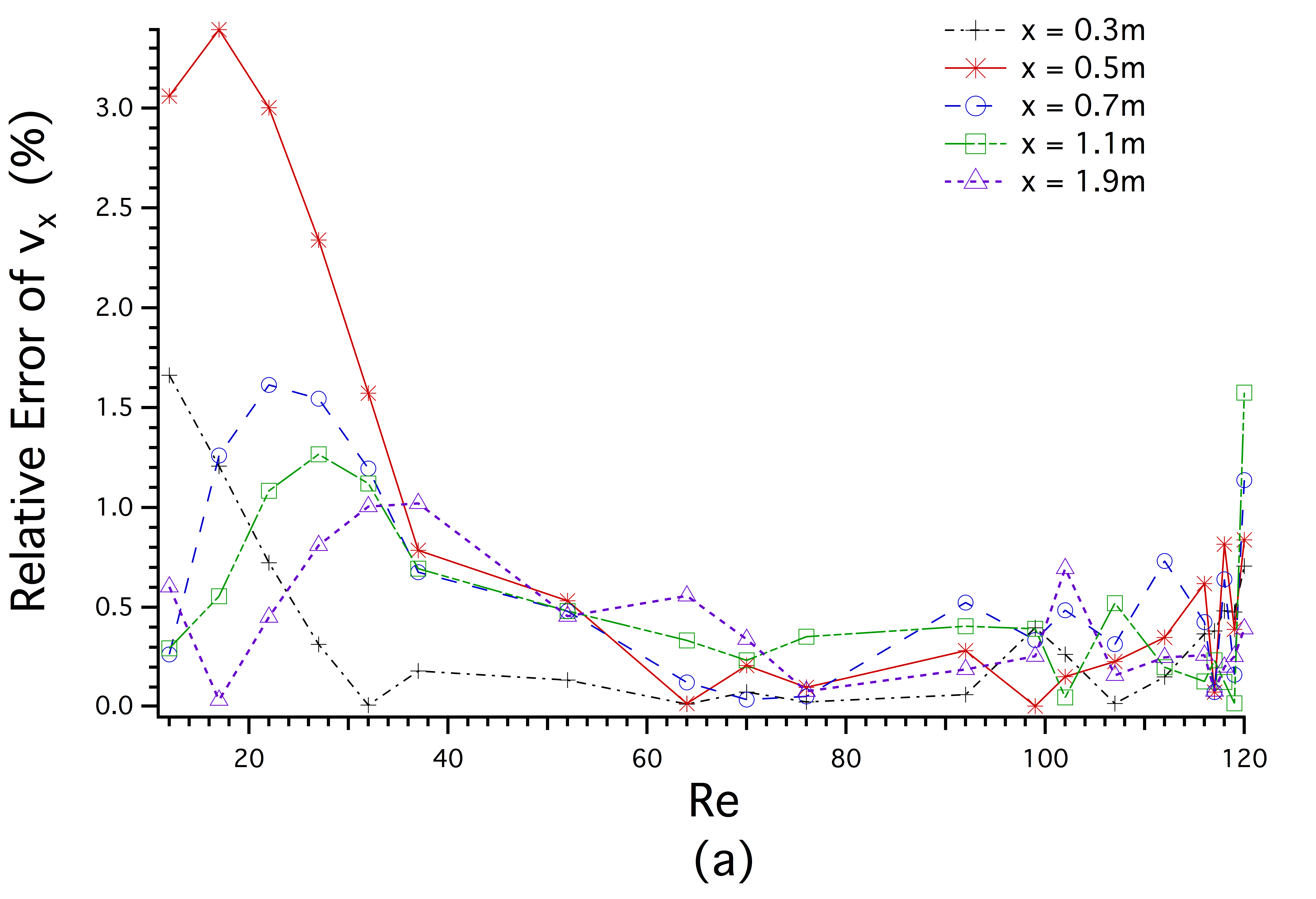}
\includegraphics[width=9.0cm]{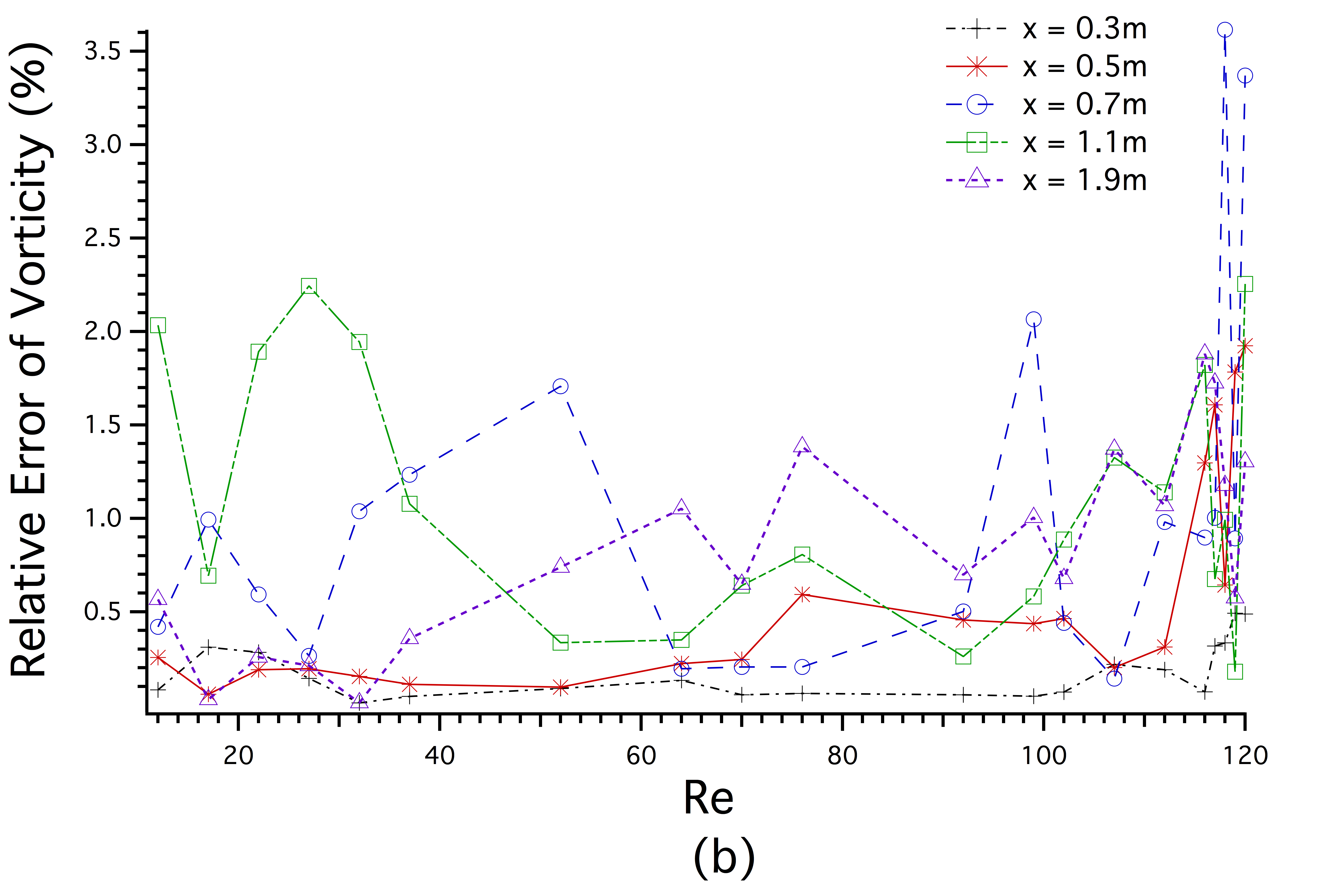}
\caption{Relative errors on predictions using 82 values of $v_x$ and the $z$-component of the vorticity of the fluid as inputs. In (a) the predictions using $v_x$ are better for fluids with a moderate $Re$ independently of the measurement location. However using the vorticity as input, the accuracy is better for low $Re$ and distances of the measurement closer to the obstacle instead of high $Re$ and farther distances (b). In both situations, the error increases in the extrapolation regime ($Re$ 115-120).}
\label{fig:RelError}
\end{figure}

The ANNs using 82 sampling points, have a relative error less than 4$\%$ in all locations, see Figure \ref{fig:RelError} for instance, where the relative errors are plotted as a function of $Re$ using $v_x$ as input for the ANN in the plot at the top and the $z$-component of the vorticity in the plot at the bottom. The results are more accurate considering $v_x$ as inputs for detectors far away from the obstacle and high values of $Re$. On the contrary, using vorticity as the input gives better results for low values of $Re$ and close distances to the cylinder decreasing accuracy at locations at 0.7m and 1.1m, perhaps due to the vortices structure. For the extrapolated cases ($Re = 115, 116, 117, 118, 119$ and $120$), the precision also decreases, but the error for those does not exceed $2\%$ and $4\%$ using $v_x$ and vorticity respectively.

On Tables \ref{tab:Predictions_Velx} and \ref{tab:Predictions_Vort}, are the averaged predictions made by the ANNs considering 82 points in the detectors, for all different $Re$ cases in all test location with their corresponding $\chi^2$ test. Considering $v_x$ as input to the network the worst adjustment is made at 0.5m, meanwhile by using the $z$-component of the vorticity the worst adjustment are at 0.7m and 1.1m. Both results are expected due to the complexity of the vortices in these regions.

\begin{table}[p]
	\begin{tabular}{|c|c|c|c|c|c|}\hline
		\multicolumn{6}{|c|}{$Re$ prediction using $v_x$} \\ \hline\hline
		{$Re$}   & 0.3m & 0.5m & 0.7m & 1.1m & 1.9m \\ \hline
		12  & 11.801  & 11.633  & 11.968  & 12.035  & 12.072  \\ \hline
		17  & 16.795  & 16.433  & 16.786  & 16.906  & 17.006  \\ \hline
		22  & 21.841  & 21.340  & 21.645  & 21.762  & 21.901  \\ \hline
		27  & 26.915  & 26.368  & 26.583  & 26.658  & 26.781  \\ \hline
		32  & 31.997  & 31.497  & 31.617  & 31.641  & 31.679  \\ \hline
		37  & 37.067  & 36.709  & 36.750  & 36.743  & 36.623  \\ \hline
		52  & 52.071  & 52.277  & 52.250  & 52.250  & 52.237  \\ \hline
		64  & 63.990  & 63.988  & 64.080  & 64.214  & 64.357  \\ \hline
		70  & 69.947  & 69.853  & 70.027  & 70.164  & 70.238  \\ \hline
		76  & 76.020  & 76.076  & 75.959  & 76.268  & 76.061  \\ \hline
		92  & 92.057  & 92.260  & 91.518  & 91.627  & 91.826  \\ \hline
		99  & 99.393  & 98.994  & 99.333  & 98.611  & 98.747  \\ \hline
		102 & 102.269 & 102.156  & 102.495 & 101.950 & 101.291 \\ \hline
		107 & 107.020 & 106.755 & 106.662 & 106.443 & 107.171 \\ \hline
		112 & 111.831 & 111.609 & 111.180 & 111.776 & 112.279 \\ \hline
		116 & 115.574 & 115.2822 & 115.506 & 116.150 & 116.302 \\ \hline
		117 & 116.553 & 116.912 & 116.912 & 116.726 & 117.095 \\ \hline
		118 & 117.430 & 117.037 & 117.244 & 118.149 & 118.238 \\ \hline
		119 & 118.432 & 118.536 & 119.193 & 119.024 & 119.303 \\ \hline
		120 & 119.152 & 118.996 & 118.636 & 118.109 & 120.471 \\ \hline \hline
		\textbf{$\chi^2$} & \textbf{0.025} & \textbf{0.103} & \textbf{0.058} & \textbf{0.054} & \textbf{0.024} \\ \hline
	\end{tabular}
	\caption{$Re$ predicted for the different detectors sampling $v_x$ on 82 points and their corresponding $\chi^2$ test, where the worst predictions were made at 0.5m from the origin. A model is considered better than other when the value of $\chi^2$ from the first model is smaller than the second one.}
	\label{tab:Predictions_Velx}
\end{table}

\begin{table}[p]
	\begin{tabular}{|c|c|c|c|c|c|}\hline
		\multicolumn{6}{|c|}{$Re$ prediction using the vorticity} \\ \hline\hline
		{$Re$}   & 0.3m & 0.5m & 0.7m & 1.1m & 1.9m \\ \hline
		12  & 11.990  & 12.031  & 12.050  & 12.244  & 12.068  \\ \hline
		17  & 17.053  & 16.990  & 17.169  & 16.882  & 16.995  \\ \hline
		22  & 22.062  & 21.958  & 22.130  & 21.584  & 21.943  \\ \hline
		27  & 27.039  & 26.947  & 26.929  & 26.394  & 26.943  \\ \hline
		32  & 32.004  & 31.951  & 31.668  & 31.378  & 32.004  \\ \hline
		37  & 36.983  & 36.959  & 36.544  & 36.601  & 37.132  \\ \hline
		52  & 51.954  & 51.950  & 52.887  & 51.826  & 52.384  \\ \hline
		64  & 63.915  & 63.858  & 64.125  & 63.777  & 64.673  \\ \hline
		70  & 69.961  & 70.172  & 70.143  & 69.552  & 70.454  \\ \hline
		76  & 75.952  & 75.549  & 75.845  & 75.387  & 77.051  \\ \hline
		92  & 92.051  & 92.420  & 92.461  & 91.760  & 92.642  \\ \hline
		99  & 99.047  & 99.432  & 101.045 & 99.576  & 99.995  \\ \hline
		102 & 101.928 & 102.472 & 101.551 & 101.096 & 101.308 \\ \hline
		107 & 107.234 & 107.215 & 106.848 & 105.582 & 108.467 \\ \hline
		112 & 112.212 & 111.651 & 110.902 & 110.724 & 113.197 \\ \hline
		116 & 115.917 & 114.496 & 114.959 & 118.113 & 113.819 \\ \hline
		117 & 116.630 & 115.119 & 115.826 & 117.790 & 119.018 \\ \hline
		118 & 117.607 & 117.241 & 113.734 & 116.829 & 116.612 \\ \hline
		119 & 118.414 & 116.878 & 117.937 & 119.214 & 119.683 \\ \hline
		120 & 119.415 & 117.692 & 115.957 & 117.295 & 118.436 \\ \hline \hline
		\textbf{$\chi^2$} & \textbf{0.001} & \textbf{0.148} & \textbf{0.406} & \textbf{0.214} & \textbf{0.197} \\ \hline
	\end{tabular}
	\caption{Same as Table \ref{tab:Predictions_Velx} but instead of $v_x$ using the $z$-component of the vorticity as input of the ANN.}
	\label{tab:Predictions_Vort}
\end{table}

\begin{figure}[p]
	\includegraphics[width=9.0cm]{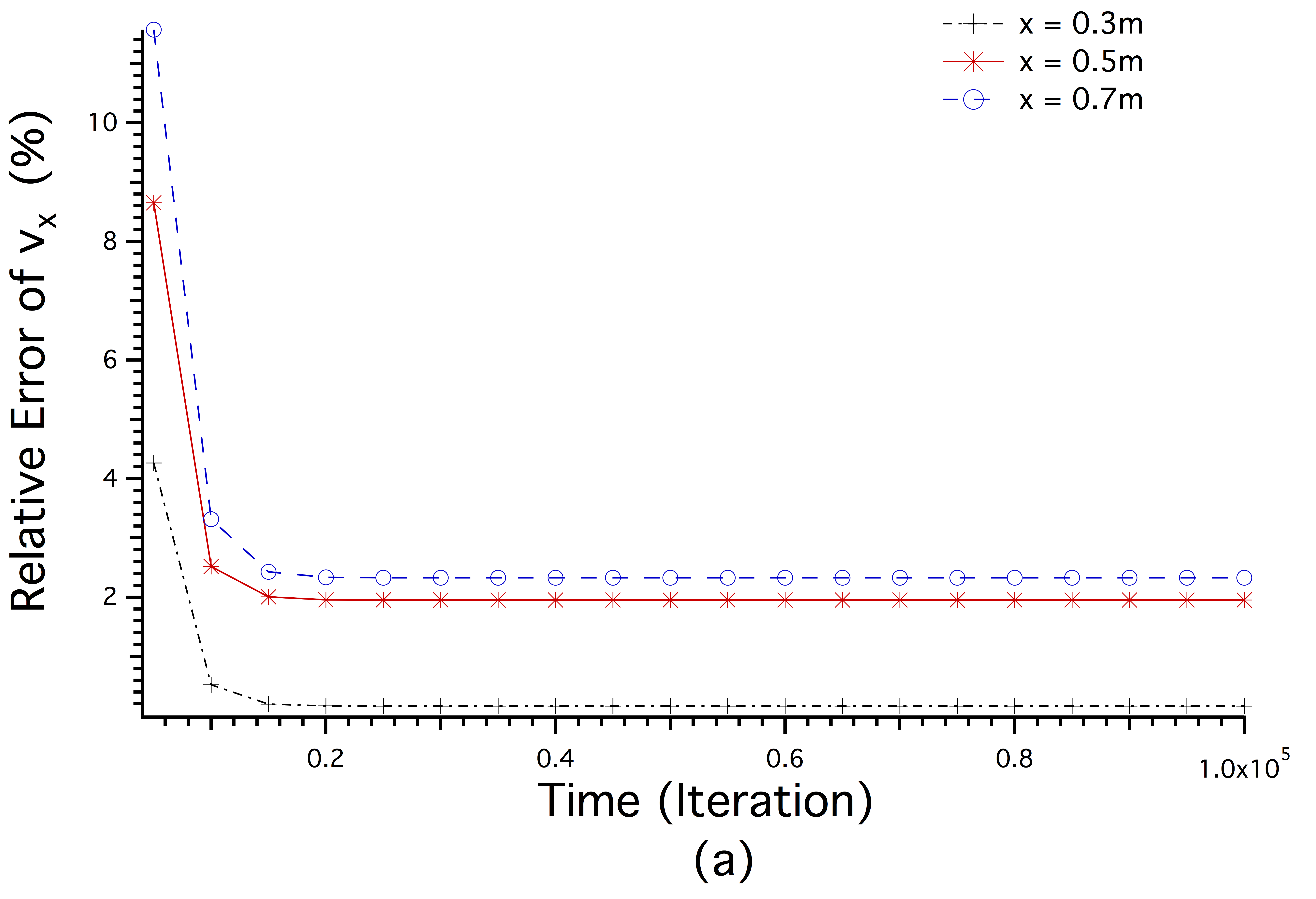}
	\includegraphics[width=9.0cm]{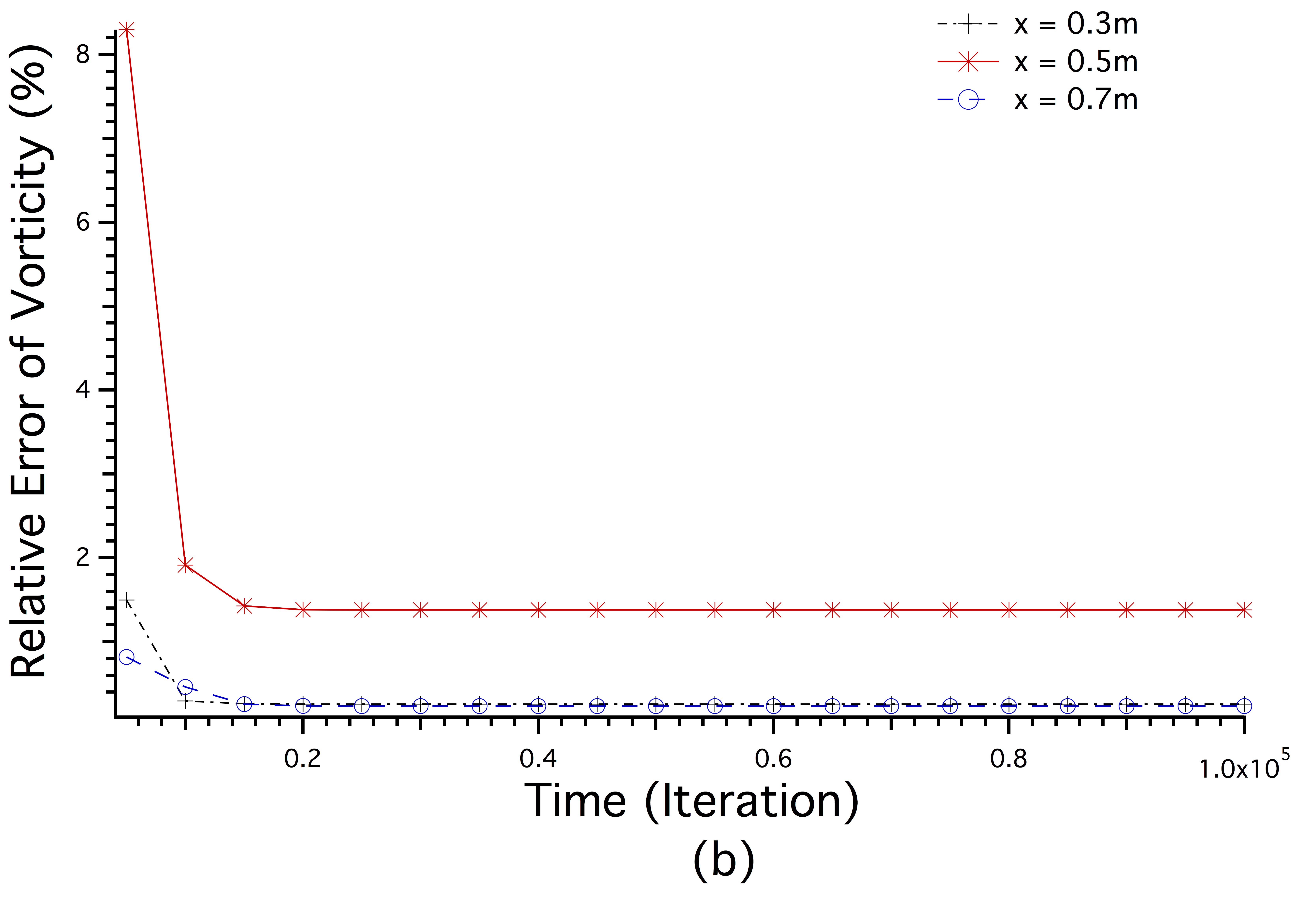}
	\caption{Relative errors on predictions using 82 values of $v_x$ (a) and the $z$-component of the vorticity (b) of the fluid as inputs for $Re=30$ along the simulation time. We observe the fluctuations on the percent error, as expected since the flow patterns change on time.}
	\label{fig:TimeRelErrorRe30}
\end{figure}

\begin{figure}[p]
	\includegraphics[width=9.0cm]{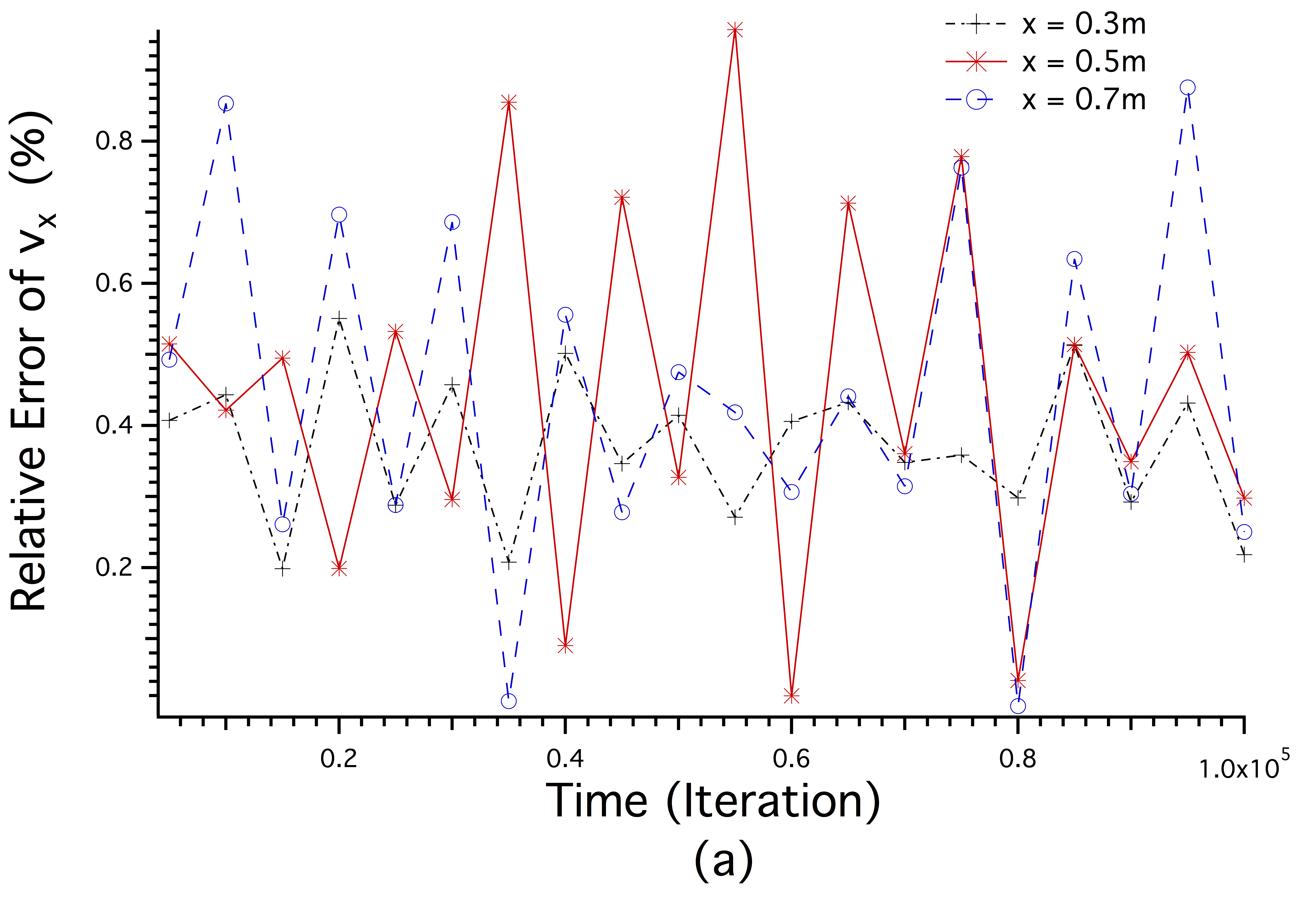}
	\includegraphics[width=9.0cm]{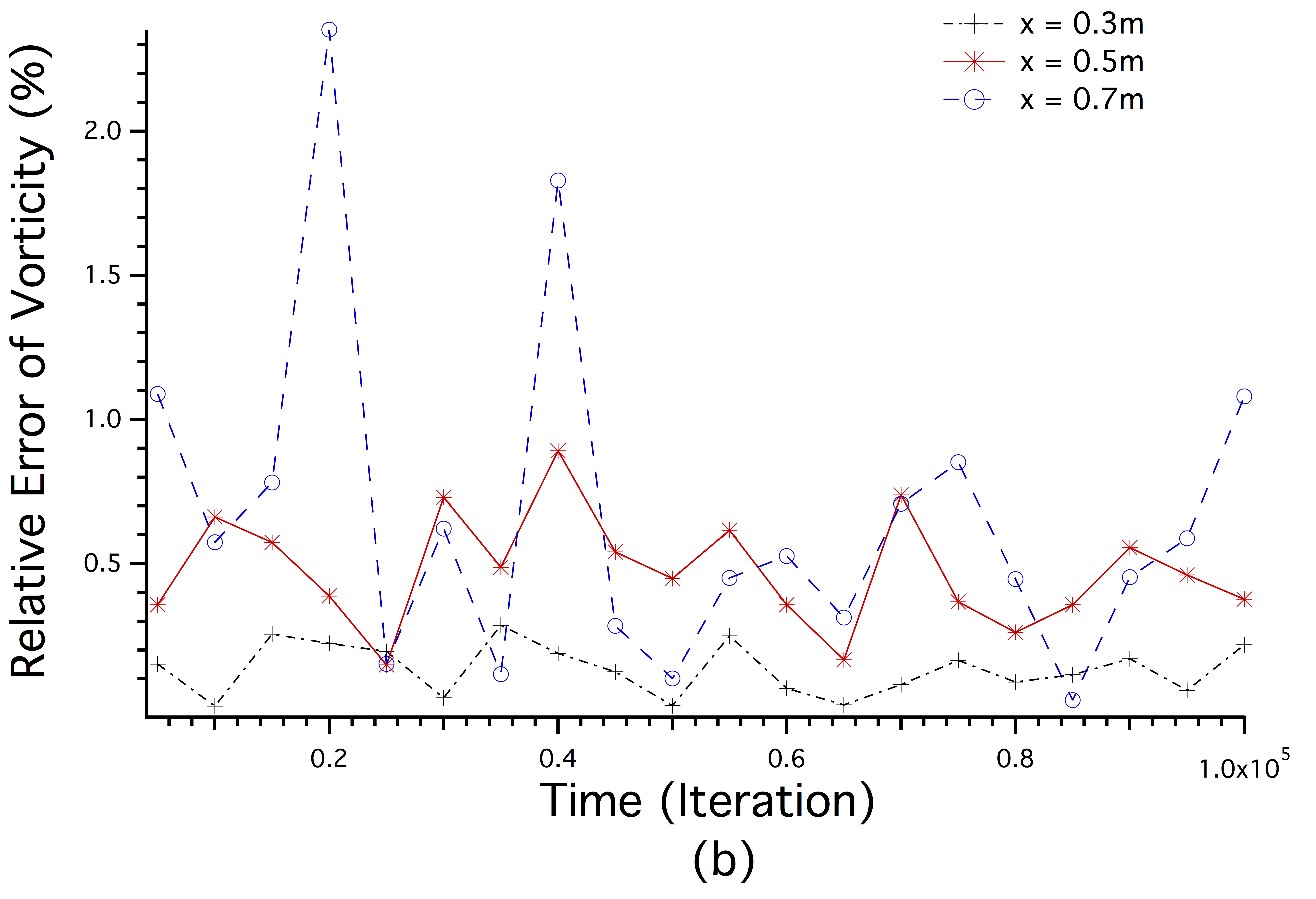}
	\caption{Relative errors on predictions using 82 values of $v_x$ (a) and the $z$-component of the vorticity (b) of the fluid as inputs for $Re=99$ along the simulation time. We observe some fluctuations on the percent error, as expected since the flow patterns change on time.}
	\label{fig:TimeRelErrorRe99}
\end{figure}

These results are obtained considering a single extraction time, and we now inspect the dependency of the ANNs predictions on different extraction times. Results are presented in Figures \ref{fig:TimeRelErrorRe30} and \ref{fig:TimeRelErrorRe99} for two different Reynolds numbers $Re = 30,99$ using the same training and validation sets. On the graphs only the first three measuring locations at 0.3m, 0.5m and 0.7m are plotted. Despite the ANNs were trained using a single time patterns, they were capable to predict within less than 5\% error for almost all cases. For $Re$=30, on the first 10000 iterations the fluid has not reached a neutral stability so the error exceeds the 5\%, afterwards the ANNs predictions converge within the expected results, as depicted on Figure \ref{fig:TimeRelErrorRe30}. In the case for $Re$ = 99, the predictions oscillate because the flow patterns also fluctuates on time as expected, maintaining the predictions on a acceptable range, Figure \ref{fig:TimeRelErrorRe99}.
  
\section{\label{sec:Conclusions} Conclusions}
From the results, we observe that using $v_x$ as input data the prediction of $Re$ is more accurate than when using the $z$-component of the vorticity in almost all cases, except for low $Re$ near the obstacle. In the extrapolation cases, the error increases as the corresponding $Re$ gets far from the set for which the ANNs were trained. The more complicated zones to predict seem to be in the middle regions, this is at 0.5m, 0.7m and 1.1m, this might be due to the complexity of the fluid's behavior. We suspect that as increasingly we modify the profile of the initial velocities, the magnitude of the resultant field of velocities will also have to increase. So the ANNs finds a clear pattern that relates the increase in the magnitude of the velocity field with the respective number of Reynolds.
Another result worth noticing is the minimum number of points over the detectors needed to obtain reliable results. In our experiments that number was 11.

An analysis over the time evolution of the flow gives a very good estimate of the possibilities to use ANNs on this kind of problems, with an expected greater error at the beginning of the simulation where fluid has not reached the neutral stability, afterwards the predictions converge. The results indicate that for general scenarios with different kind of blocks the dependence in time and space could be more complicated, and the averages in space and time of the velocity profiles have to be taken in account. It is expected that the performance could be increased if they are trained with the flow patterns over time and using a more complex ANN structure like a recurrent neural network. 

We can conclude that the method presented in this paper is strong enough to estimate the values of the Reynolds number measuring the profile of velocities or vorticity of the fluid. This approach can be used to obtain the initial parameters or inputs used on the simulation like the diameter of the obstacle, initial velocity or other physical properties characterizing the problem. With this in mind, we look towards a better implementation of a machine learning algorithm capable to characterize, among other interesting topics, blocked flows in a pipe.

\begin{acknowledgments}
This research is partially supported by grant CIC-UMSNH-4.23. The authors would like to thank Francisco S. Guzm\'an for his comments after reading the manuscript.
\end{acknowledgments}


\end{document}